\documentclass{jps-cp-overleaf}
\usepackage{txfonts} 

\title{Multiple Superconducting Phases under Pressure for Different Field Directions in UTe$_2$}

\author{
Dai \textsc{Aoki}$^{1}$,
Atsushi \textsc{Miyake}$^{1}$,
Ai \textsc{Nakamura}$^{1}$,
Yoshiya \textsc{Homma}$^{1}$,
Dexin \textsc{Li}$^{1}$,
Shota \textsc{Mizuno}$^{1}$,
Georg \textsc{Knebel}$^{2}$,
Daniel \textsc{Braithwaite}$^{2}$, 
Jean-Pascal \textsc{Brison}$^{2}$, and 
Jacques \textsc{Flouquet}$^{1,2}$
}

\inst{
$^{1}$Institute for Materials Research, Tohoku University, Oarai, Ibaraki 311-1313, Japan  \\
$^{2}$Univ. Grenoble Alpes, CEA, Grenoble INP, IRIG, PHELIQS, F-38000 Grenoble, France
}

\email{aoki@imr.tohoku.ac.jp}

\recdate{September 24, 2026}
\abst{
The existence of multiple superconducting phases is a striking feature of UTe$_2$, arising from the spin and orbital degrees of freedom and supporting a spin-triplet superconducting state.
We present field--temperature ($H$--$T$) phase diagrams under pressure for fields applied along the three principal crystallographic axes, as determined by AC calorimetry measurements.
The pressure evolution of the multiple superconducting phases is highly anisotropic with respect to magnetic field direction, revealing rich superconducting phenomena arising from complex fluctuations.
}

\kword{UTe$_2$, spin-triplet state, superconductivity, high pressure}

\begin{document}
\maketitle

UTe$_2$ has attracted much attention because of its unusual superconducting properties~\cite{Ran19,Aok19_UTe2,Aok22_UTe2_review}. 
One of the highlights observed in UTe$_2$ is the field-reentrant superconductivity. 
When the field is applied along the hard-magnetization axis, the $b$ axis, superconductivity survives up to the metamagnetic transition field, $\mu_0 H_{\rm m}\sim 34\,{\rm T}$, exhibiting field-reentrant behavior~\cite{Kne19}, 
although the superconducting transition temperature is only $T_{\rm c}=1.6$--$2.1\,{\rm K}$.
This robustness of superconductivity against magnetic field far exceeding the Pauli limit is similar to that observed in ferromagnetic superconductors, such as URhGe and UCoGe. 
Soon after the discovery of superconductivity in UTe$_2$, it was proposed that UTe$_2$ is an endmember of ferromagnetic superconductors; that is, UTe$_2$ is a heavy fermion paramagnet on the verge of ferromagnetic order. 
This is a good starting point to understand UTe$_2$; however, it is oversimplified.
In fact, antiferromagnetic fluctuations with an incommensurate wavevector are detected in the inelastic neutron scattering experiments~\cite{Dua20,Kna21_PRB}. 
Furthermore, above the critical pressure, $P_{\rm c}\sim 1.45\,{\rm GPa}$, at which superconductivity is suppressed, the antiferromagnetic order is observed in the magnetic susceptibility and neutron scattering experiments~\cite{Li21,Kna25}.
Thus, UTe$_2$ is not simply an endmember of ferromagnetic superconductors. 

\begin{figure}[tbh]
\begin{center}
\includegraphics[width= 0.9\hsize]{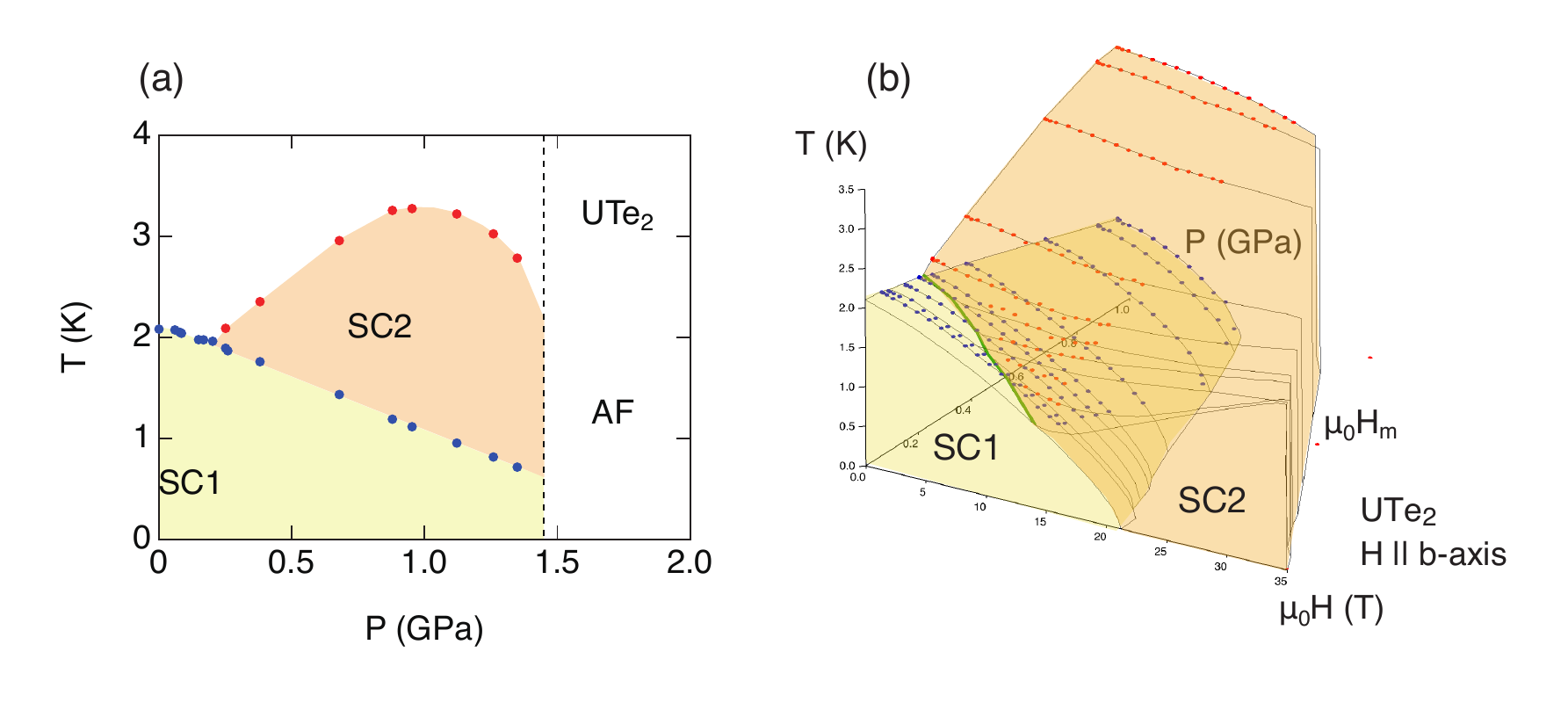}
\caption{(a) $T$--$P$ phase diagram at zero field in UTe$_2$ determined by the AC calorimetry measurements using a high-quality sample ($T_{\rm c}=2.08\,{\rm K}$ defined by zero resistivity, the residual resistivity ratio, $\mbox{RRR}\sim 500$, the scaled residual Sommerfeld coefficient, $\gamma_0/\gamma_{\rm N} = 0.03$).
Above the critical pressure ($P_{\rm c}\sim 1.45 \,{\rm GPa}$)~\cite{Aok21_UTe2}, the antiferromagnetic state (AF) is stabilized~\cite{Li21,Kna25}.
(b) Semi-schematic $T$--$P$--$H$ phase diagram for $H \parallel b$-axis.
}
\label{fig:fig_TPH}
\end{center}
\end{figure}

Under pressure, superconductivity of UTe$_2$ shows rich phenomena. 
As shown in Fig.~\ref{fig:fig_TPH}(a), multiple superconducting phases associated with the splitting of $T_{\rm c}$ were first reported in the AC calorimetry measurements at zero field~\cite{Bra19}.
With increasing pressure, $T_{\rm c}$ slightly decreases and starts splitting around $0.2\,{\rm GPa}$.
The lower transition temperature decreases continuously with increasing pressure, whereas the upper transition temperature increases to a maximum value, $T_{\rm c}\sim 3\,{\rm K}$ at around $1\,{\rm GPa}$, and then slightly decreases with further increasing pressure. 
Note that recent ultrasound measurements have revealed an additional phase boundary within the SC1 region near $0.2\,{\rm GPa}$,~\cite{Kam26} which we have not detected in our AC calorimetry measurements.

Above $P_{\rm c}$, superconductivity is suppressed at zero field, and the antiferromagnetic state is stabilized. 
Interestingly, at pressures slightly above $P_{\rm c}$, superconductivity is observed at high fields for $H\parallel c$-axis, while no superconductivity appears at zero field, revealing field-induced superconductivity~\cite{Aok21_UTe2}.
Superconductivity emerges at high fields where the antiferromagnetic state is suppressed. 
Thus, this superconducting phase seems to compete with the pressure-induced antiferromagnetic state.

The link between field-reentrant superconductivity and multiple superconducting phases has recently been clarified.
As mentioned, field-reentrant superconductivity appears for $H\parallel b$-axis at ambient pressure. 
In the field-reentrant superconducting phase, high-field specific heat measurements clearly detected a phase boundary that extends continuously into the field-reentrant region,
indicating the multiple superconducting phases even at ambient pressure~\cite{Rosuel23}.
As the pressure is increased slightly, the high-field superconducting phase gradually shifts to the low-field region, 
and finally $T_{\rm c}$ splits at zero field above $\sim 0.2\,{\rm GPa}$~\cite{Vas25}.
These measurements establish a continuous connection between the high-field superconducting phase at ambient pressure and the high-temperature superconducting phase (SC2) at zero field under pressure, demonstrating a clear link between field-reentrant superconductivity and multiple superconducting phases.
As semi-schematically shown in Fig.~\ref{fig:fig_TPH}(b), the SC1 phase is embedded within the larger SC2 region in the $T$--$P$--$H$ phase diagram. 
Different superconducting order parameters and symmetries are expected. 
The identification of these phases, however, remains under debate.

In order to clarify the multiple superconducting phases in more detail, we performed  AC calorimetry measurements at temperatures down to $0.1\,{\rm K}$ and at high fields up to $15\,{\rm T}$. 
Single crystals of UTe$_2$ were grown using the chemical vapor transport (CVT) method. 
The details are described in Ref.~\citen{Aok24_UTe2_CrystalGrowth}.
On the other hand, the data shown in Fig.~\ref{fig:fig_TPH}(a) were obtained using high-quality single crystals obtained by the molten salt flux liquid transport (MSFLT) method~\cite{Aok24_UTe2_CrystalGrowth}.
The AC calorimetry measurements were performed using the Au-Au/Fe thermocouple and Au wires as a heater.
The small, oriented single crystals spot-welded to the thermocouple and the heater were inserted into a hybrid-type piston cylinder cell. 
The pressure-transmitting medium was Daphne 7373, and the pressure was determined by the superconducting transition of Sn.

\begin{figure}[tbh]
\includegraphics[width= 1\hsize]{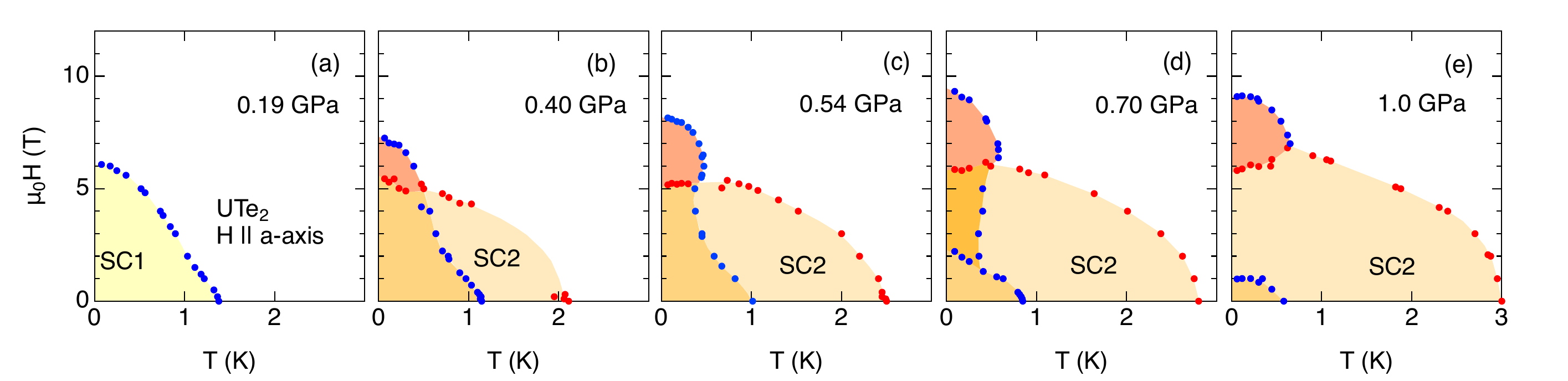}
\caption{$H$--$T$ phase diagrams for $H\parallel a$-axis at different pressures in UTe$_2$. The data points are taken from our previous AC calorimetry measurements.~\cite{Aok20_UTe2}}
\label{fig:fig_HT_a}
\end{figure}

Figure~\ref{fig:fig_HT_a} shows the pressure evolution of the $H$--$T$ phase diagram for the field along the $a$-axis, corresponding to the easy-magnetization axis~\cite{Aok20_UTe2}.
$T_{\rm c}$ at ambient pressure is $1.6\,{\rm K}$, which is lower than that of the high-quality sample with $T_{\rm c}\sim 2.1\,{\rm K}$, 
but the overall behavior of the $H$--$T$ phase diagram for the $T_{\rm c}=1.6\,{\rm K}$ sample is likely the same as that for the $T_{\rm c}\sim 2.1\,{\rm K}$ sample.
At $0.4\,{\rm GPa}$, $H_{\rm c2}$ shows a sudden increase on cooling.
This is clearer at $0.54\,{\rm GPa}$.
The $H_{\rm c2}$ curve associated with the SC2 phase with $T_{\rm c}=2.5\,{\rm K}$ at zero field shows convex curvature,
while the internal superconducting phase boundary becomes nearly vertical and joins the $H_{\rm c2}$ curve at its sharp upturn.
It is clear that this unusual $H_{\rm c2}$ enhancement at low temperatures is due to the multiple superconducting phases.
At higher pressure, the high-field superconducting phase starts to decouple from the low-field superconducting phase,
as shown in Figs.~\ref{fig:fig_HT_a}(d) and \ref{fig:fig_HT_a}(e).

\begin{figure}[tbh]
\includegraphics[width= 1\hsize]{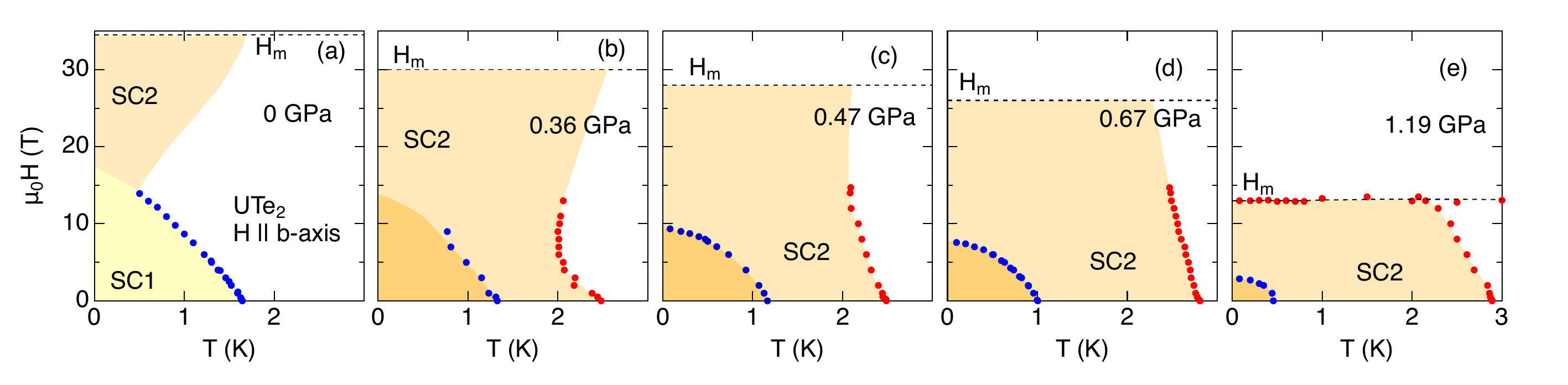}
\caption{$H$--$T$ phase diagrams for $H\parallel b$-axis at different pressures in UTe$_2$. The dotted lines indicate the metamagnetic transition field $H_{\rm m}$, estimated from Refs.~\citen{Kne20,Lin20,Miy22}. The lines are guides to the eye.}
\label{fig:fig_HT_b}
\end{figure}

When the field is applied along the $b$-axis, the evolution of the $H$--$T$ phase diagram is very different.
As shown in Fig.~\ref{fig:fig_HT_b}, the reentrant superconductivity appears above $15\,{\rm T}$ and continues up to the first-order metamagnetic transition field $\mu_0 H_{\rm m}\sim 34\,{\rm T}$~\cite{Kne19,Rosuel23}.
Superconductivity is abruptly destroyed above $H_{\rm m}$ due to the suppression of the magnetic fluctuations and/or a drastic change of the electronic states.
The reentrant superconducting phase is the SC2 phase, which is identical to the high-pressure and high-temperature superconducting phase described earlier.
$H_{\rm m}$ decreases with increasing pressure~\cite{Kne20,Lin20}, as indicated by dotted lines in Fig.~\ref{fig:fig_HT_b}.
As superconductivity is suppressed above $H_{\rm m}$, the reentrant behavior is weakened, and the magnitude of the initial slope of $H_{\rm c2}$ increases.
Notably, the low-temperature SC1 phase shrinks with pressure, and its phase boundary shows convex curvature. 
This behavior again contrasts markedly with that for $H\parallel a$-axis. 

\begin{figure}[tbh]
\includegraphics[width= 1\hsize]{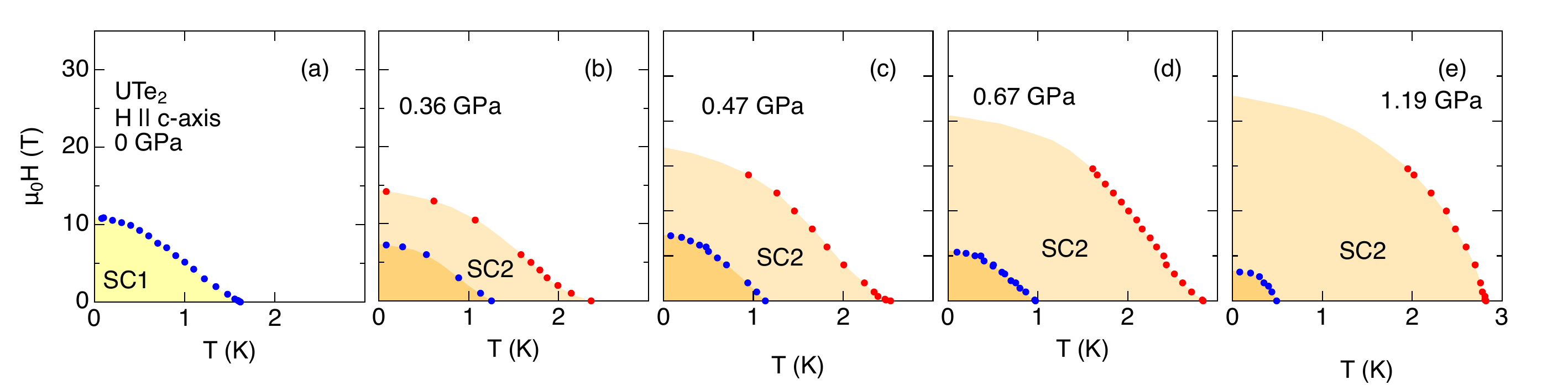}
\caption{$H$--$T$ phase diagrams for $H\parallel c$-axis at different pressures in UTe$_2$. The lines are guides to the eye.}
\label{fig:fig_HT_c}
\end{figure}

Figure~\ref{fig:fig_HT_c} shows the $H$--$T$ phase diagrams at different pressures for $H\parallel c$-axis.
After $T_{\rm c}$ splits, the low-temperature transition lines associated with SC1 progressively contract with increasing pressure, as do those for $H\parallel b$-axis.
The $H_{\rm c2}$ curves for the SC2 phase display a larger magnitude of the initial slope with increasing pressure, as also observed for $H\parallel b$-axis.
Unlike the case for $H \parallel b$-axis, no metamagnetic transition is observed for $H\parallel c$-axis.
The SC2 phase appears to extend beyond $15\,{\rm T}$.
A very large $H_{\rm c2}$ and a nearly vertical $H_{\rm c2}(T)$ curve near $T_{\rm c}$ have indeed been observed at pressures slightly lower than $P_{\rm c}$~\cite{Aok21_UTe2}.

At ambient pressure, $H_{\rm c2}$ at low temperatures is already quite anisotropic, including the field-reentrant superconductivity.
The angular dependence of $H_{\rm c2}$ cannot be described by the effective-mass model, because of the anisotropic magnetic fluctuations and the field response~\cite{Aok24_UTe2_HighField}.
On the other hand, the angular dependence of the initial slope of $H_{\rm c2}$ determined by field- and angle-dependent AC calorimetry measurements can be fitted approximately using the effective-mass model, 
which is consistent with the quasi-2D Fermi surfaces detected by the de Haas-van Alphen effect~\cite{Aok23_UTe2_dHvA}.

Under pressure, the $H$--$T$ phase diagrams for different field directions are quite complicated, as already shown. 
The field response of the multiple superconducting phases is quite anisotropic, revealing the anisotropic field response of the magnetic fluctuations as well as the change of the electronic states.

\begin{figure}[tbh]
\includegraphics[width= 1\hsize]{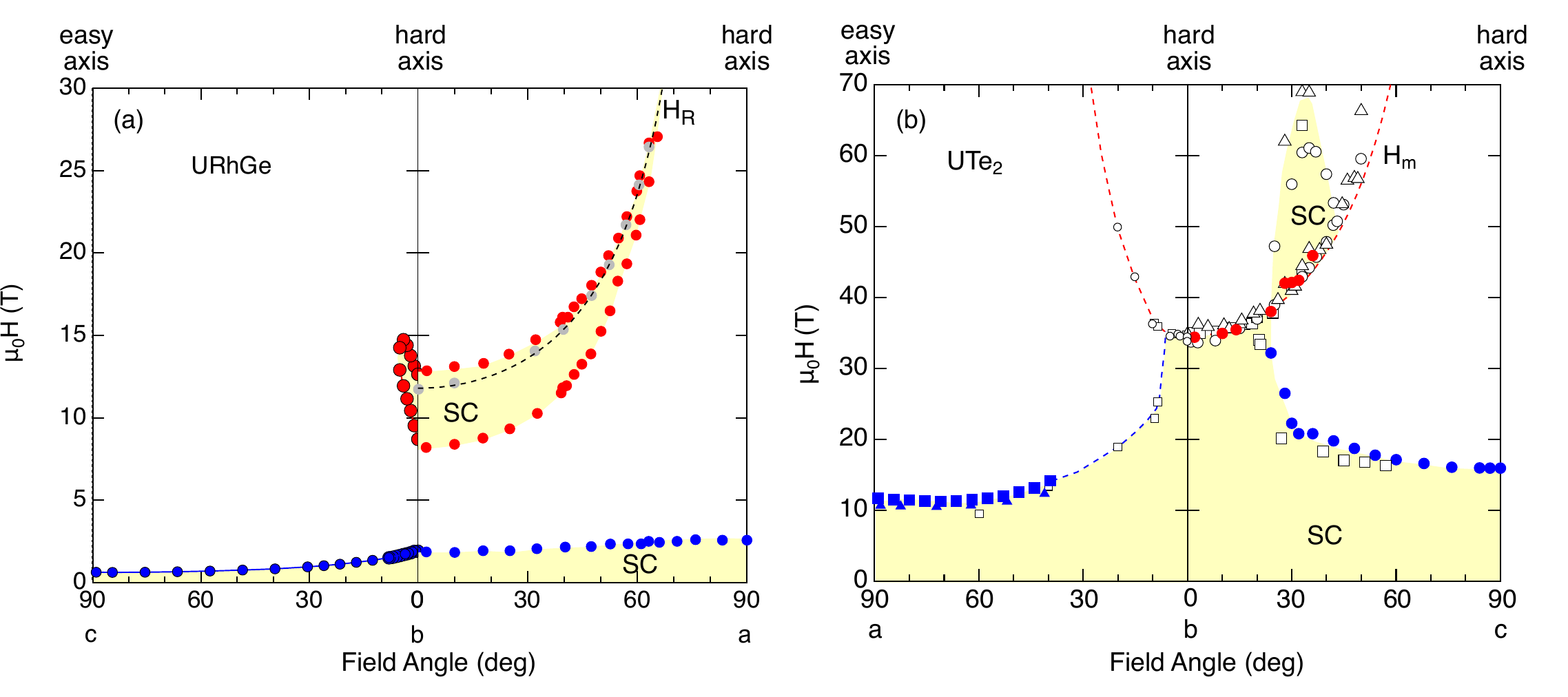}
\caption{Angular dependence of the superconducting critical fields in (a) URhGe and (b) UTe$_2$. SC, $H_{\rm R}$ and $H_{\rm m}$ denote superconductivity, the spin-reorientation field and the metamagnetic transition field, respectively. In URhGe, the easy-magnetization axis is the $c$-axis, and the hard axes correspond to the $b$ and $a$-axes. In UTe$_2$, the magnetization easy axis is the $a$-axis, and the hard axes correspond to the $b$ and $c$ axes. The high-field data in URhGe are taken from Ref.~\citen{Lev07}. The data of UTe$_2$ are taken from Refs.~\citen{Aok24_UTe2_HighField,Wu24,Ran19_HighField}.}
\label{fig:fig_URhGe_UTe2}
\end{figure}

The easy-magnetization axis at ambient pressure corresponds to the $a$-axis, and the hardest magnetization axis is the $b$-axis in UTe$_2$.
As shown in Fig.~\ref{fig:fig_URhGe_UTe2}(b), field-reentrant superconductivity is favored for fields along hard-magnetization axes.
The angular dependence of $H_{\rm c2}$ and of the boundaries of the field-reentrant phase are qualitatively similar to those observed in ferromagnetic superconductors, URhGe and UCoGe. 
For instance, as shown in Fig.~\ref{fig:fig_URhGe_UTe2}(a), the reentrant superconductivity in URhGe survives when the field direction is kept along the hard-magnetization axis~\cite{Lev07}, while tilting the field toward the easy-magnetization axis destroys superconductivity immediately.
This is because the ferromagnetic fluctuations are suppressed by tilting the field direction to the easy-magnetization axis. 

Similar trends are also observed in UTe$_2$. 
When the field is tilted from the $b$-axis toward the $c$-axis (from one hard-magnetization axis toward another), reentrant superconductivity below $H_{\rm m}$ survives up to around $20\,{\rm deg}$ and then superconductivity appears above $H_{\rm m}$~\cite{Ran19_HighField,Wu24,Aok24_UTe2_HighField,Mep26}.
Superconductivity is favored for fields along the hard-magnetization axes, but the mechanism seems to be more complicated than that for ferromagnetic superconductors.

In summary, we performed the AC calorimetry measurements under pressure with the field applied along the $a$, $b$, and $c$-axes. 
The pressure evolution of the $H$--$T$ phase diagrams differs markedly among the three field directions. 
These results provide constraints on the roles of anisotropic magnetic fluctuations and possible Fermi-surface changes in stabilizing the multiple superconducting phases.

\section*{Acknowledgements}
This work was supported by KAKENHI (JP22H04933, JP24H01641, JP25H01249), and ICC-IMR.


\end{document}